\documentclass[]{elsart}

\usepackage{epsf}
\usepackage{psfrag}

\begin{document}
\begin{frontmatter}

\title {Analysis of telephone network traffic based on a complex user network}
\author{Yongxiang Xia\corauthref{cor}}\ead{enyxxia@eie.polyu.edu.hk}, \author{Chi K.~Tse}, \author{Francis C.~M.~Lau}, \author{Wai Man Tam}, \author{Michael Small}
\corauth[cor]{Corresponding author. Tel.: (852) 2766 4745; fax: (852) 2362 8439.}
\address{Department of Electronic and Information Engineering, Hong Kong Polytechnic University, Hong Kong, China}

\begin{abstract}
The traffic in telephone networks is analyzed in this paper. Unlike the classical traffic analysis
where call blockings are due to the limited channel capacity, we consider here a more realistic
cause for call blockings which is due to the way in which users are networked in a real-life human
society. Furthermore, two kinds of user network, namely, the fully-connected user network and the
scale-free network, are employed to model the way in which telephone users are connected. We show
that the blocking probability is generally higher in the case of the scale-free user network, and
that the carried traffic intensity is practically limited not only by the network capacity but
also by the property of the user network.
\end{abstract}

\begin{keyword}
Telephone network, traffic analysis, complex networks, scale-free network.

\PACS 89.75.Hc, 89.75.Da, 89.75.Fb
\end{keyword}
\end{frontmatter}


\section{Introduction}

Telephone systems have undergone rapid developments in the past few decades. With a
growing number of end users and an increasing demand for greater variety of services,
operators are facing challenges in providing a variety of communication services and at the
same time maintaining an adequate level of  quality of service \cite{garcia,Anurag}.
As the system has a limited capacity,
it can only support a limited amount of simultaneous traffic. Under this condition, traffic analysis
that can reflect the true system behavior is indispensable. In the past few decades, several
traffic models for communication systems have been proposed \cite{bellamy}. They are derived by
fitting the existing traffic data under particular sets of conditions. However, since the
underlying mechanisms of the traffic behavior are unknown or simply not taken into account in the
modeling process, such models fall short of a clear connection with the actual physical processes
that are responsible for the behavior observed in the traffic data.

Recent studies on the so-called \textit{complex networks} \cite{watts,albert,newman} provide a
novel perspective for understanding the traffic loads in some communication systems
\cite{zhao,ohkubo,sole,lee}.  The communication networks studied are
packet-switching networks, in which the information packet goes through nodes between the source
and the destination. During the transmission process no dedicated connection is set up. As a
result, packets from different sources may share a communication link in the system. On the
other hand, telephone networks are {\em circuit-switching} systems, in which a dedicated channel is
occupied for each telephone call. This channel cannot be used by another user during the call
holding time. Because the mechanisms are different, the analysis applied to the study of
packet-switching networks are not applicable to the telephone networks.

Aiello \textit{et al.} \cite{aiello} studied the daily traffic of long-distance calls in a
telephone network and found that the incoming and outgoing connections in this network follow a
power-law distribution. This work represents the earliest attempt to understand the traffic performance of
telephone systems. However, the approach was based on fitting the existing traffic data, and the underlying mechanisms are still unexplored.

Obviously, the traffic performance of a telephone
system is strongly correlated with the behavior of its users. From the viewpoint of complex
networks, the users in a telephone system form a complex user network. This user network is a
social network, which exhibits small-world and scale-free properties \cite{albert}. In our
previous work \cite{Xia}, a scale-free \textit{user network model}\/ was used in the analysis of
telephone network traffic, which has resulted in  a clear connection between the user network
behavior and the system traffic load. Specifically, we assumed that the telephone system had an
infinite capacity so
that we might focus our attention on the effect of user network behavior on the system traffic load.
However, such an assumption was ideal since no real communication system has an infinite capacity.

In this paper we consider a more complex but realistic scenario, in which both the limited capacity
and user engagement may cause call blockings and thus influence the system traffic load.
We illustrate how network traffic load can be more realistically simulated, and show that telephone
traffic (including mobile network traffic) should be considered by taking into account both the system
capacity and the user network behavior.

\section{Traffic analysis in telephone networks}
\label{Traffic_Analysis}

In a telephone network, ``traffic'' refers to the accumulated number of communication channels
occupied by all users. Different from the user network, the telephone network is a directed
complex network, in which each edge has a direction from the caller to the receiver. For each
user, the call arrivals can be divided into two categories: \textit{incoming calls} and
\textit{outgoing calls}. Here, incoming calls refer to those being received by a user, and
outgoing calls refer to those being initiated by that user. Since every incoming call for one user
must be originated from an outgoing call of another user, we only need to consider outgoing calls
from each user when we analyze the network traffic. If not specified, the term {\em call arrival}
means {\em outgoing call arrival} in the rest of this paper.

\begin{figure}[t]
\centerline{
\epsfxsize=8cm\epsfbox{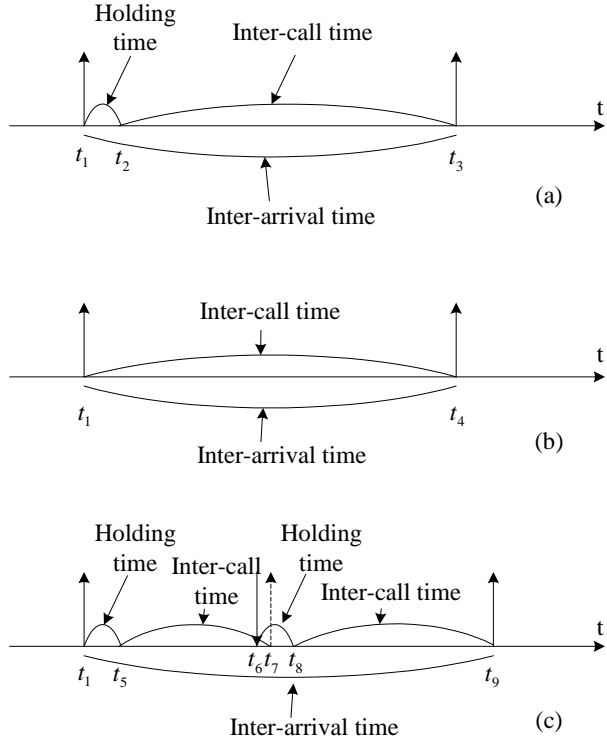}}
\caption{\label{fig:time} Three typical calling processes.}
\end{figure}

Outgoing calls are initiated randomly. If a call arrives and the conversation is successfully
established, both the caller and the receiver will be engaged for a certain duration commonly
known as \textit{holding time}. The length of the holding time is also a random variable. Thus,
the traffic load depends on the rate of call arrivals and the holding time for each call.
Figure \ref{fig:time} shows three cases of the calling process.

\noindent \textit{Case I: Call established.} When an outgoing call arrives at time
$t_1$, a receiver is randomly selected. If this receiver is idle at that time and the network has an
idling channel, a call is successfully established. The caller and receiver will be engaged for
a duration of holding time $(t_2-t_1)$. The call ends at time $t_2$. The \textit{inter-call time}
$(t_3-t_2)$ is the duration between the end of this call and the beginning of the next outgoing
call arrival. Also, the inter-arrival time is equal to the sum of the holding time and the
inter-call time. This is the normal calling process and is depicted in Fig.~\ref{fig:time} (a).

\noindent \textit{Case II: Call blocked.} Suppose two outgoing calls are made at $t_1$ and $t_4$.
A call blocking may occur due to two reasons. First,
if the receiver is engaged with another call at time $t_1$, then any new call attempt will fail
and a call blocking is said to occur. Another reason for call blockings is the limited network
capacity. If all channels are occupied at time $t_1$, the new call attempt is blocked.
The telephone network is usually considered as a ``lossy'' system, in which the blocked call
simply ``disappears''  from the network. In this case the inter-arrival time is equal to the inter-call
time (i.e., $t_4-t_1$), as shown in Fig.~\ref{fig:time} (b).

\noindent \textit{Case III: Call cancelled.} In this case, an outgoing call is supposed to take
place at time $t_7$. However, if an incoming call has arrived ahead of it and the conversation is
still going on at time $t_7$, the outgoing call attempt will be cancelled. Since this call attempt
has not been initiated, it is counted as neither  a call arrival nor a call blocking. When the
conversation ends at time $t_8$, another inter-call time is assumed before the next outgoing call
arrives at time $t_9$. In this case, the inter-arrival time is $(t_9-t_1)$, as illustrated in
Fig.~\ref{fig:time} (c). Of course, at time $t_9$, it is also possible that the user is engaged
with another call. Then, the call arrival at time $t_9$ will again be cancelled, and the
inter-arrival time will be extended accordingly.

Clearly, the shortest inter-arrival time is equal to the inter-call time, which
happens only in Case II. In our subsequent analysis, the above three cases of the call arrival
process will be considered. Here, we note that in some previous study, simplifying assumptions are
made about this process leading to drastic simplification of the analysis \cite{Siemens,Thompson}.
However, we prefer to study the traffic without making any simplifying assumptions on the call
arrival process to avoid obscuring the effects of the small-world scale-free networks.

The holding time and the inter-call time are usually modelled by some random variables with
exponential distribution. The probability density function (PDF) of the holding time is given by
\begin{equation}
f_1(t)=\frac{1}{t_m} e^{-t/t_m},
\end{equation}
where $t_m$ is the average holding time. The PDF of the inter-call time is given by
\begin{equation}
f_2(t)=\mu_i e^{-\mu_i t},
\end{equation}
where $1/\mu_i$ is the average inter-call time. The holding times are distributed identically for
all users, but the mean values of the inter-call times for different users may be different.

As shown in Fig.~\ref{fig:time}, the inter-arrival times for the three cases are different.
However, if we examine the traffic over a sufficiently long period of time (e.g., 60 min), we can
obtain the average call arrival rate $\lambda_i$, which is the average number of call arrivals per
unit time, for user $i$. Thus, the average arrival rate for the whole network is
\begin{equation}
\lambda=\sum_{i=1}^N {\lambda_i}  \label{eq:lambda},
\end{equation}
where $N$ is the total number of users in the network.

A commonly used measure of traffic is the
\textit{traffic intensity} \cite{bellamy}, which is defined by
\begin{equation}
A=2\lambda t_m, \label{eq:erlang}
\end{equation}
and represents the average traffic offered over a period of time. It is
dimensionless, but is customarily expressed in units of erlang. It should be noted that there is
a coefficient ``2" in the above equation. This is because in our study we assume that both the caller
and the receiver stay in the same telephone network. Thus, two channels are used for each call conversation.

In a telephone network, there are two distinct kinds of traffic:  \textit{offered traffic} and
\textit{carried traffic.} The offered traffic is the total traffic that is being requested by
users. The carried traffic is the actual traffic that is being carried by the network. It can be
found as the sum of the holding times of all call conversations. In
practice, due to limited network capacity and some user behavior,
a certain percentage of the offered traffic experiences {\em network
blocking.} Hence, the carried traffic is smaller than the offered traffic.
The carried traffic, moreover, can be expressed by
\begin{equation}
A_{\rm carried} = A_{\rm offered}  (1-p_{\rm blocking} ) = 2\lambda t_m (1-p_{\rm
blocking}), \label{eq:carried-traffic}
\end{equation}
where $A_{\rm carried}$ and $A_{\rm offered}$ denote the carried traffic
and the offered traffic, respectively, and $p_{\rm blocking}$ represents the blocking
probability of a call.

The telephone network is typically measured in terms of the average activity during the busiest
hour of a day \cite{bellamy}. During the busiest hour, the average contribution of one user to the
traffic load is typically between 0.05 and 0.1 erlang. The average holding time is 3 to 4 min.

\section{User network configuration}

Formally, we may describe a user network in terms of nodes and connections. A node is a user, and
a connection between two nodes indicates a possibility that these two users may call each other,
i.e., a connection connects a pair of acquaintances.

In the classical traffic analysis, each user can call any other user with equal probability. Thus,
the user network is a fully-connected network. In such a user network, the effect of each user is
assumed to be identical.
However, in reality, some users make more calls than others do. A relatively small group of users
are usually responsible for most of the calls and hence are having a comparatively bigger impact
to the traffic. Our basic assumption of the user network is that it is not uniform, i.e., a user
does not call every user in the network with equal probability. In fact, users usually only call
their own acquaintances, such as family members, colleagues and friends. If a user has more
acquaintances, the probability of him making/receiving a call at any time is higher. Thus, in the
real user network, user $i$ only has $n_i$ connections which connect him to his $n_i$
acquaintances.

\begin{figure}[!t]
\centerline{\epsfxsize=8.75cm\epsfbox{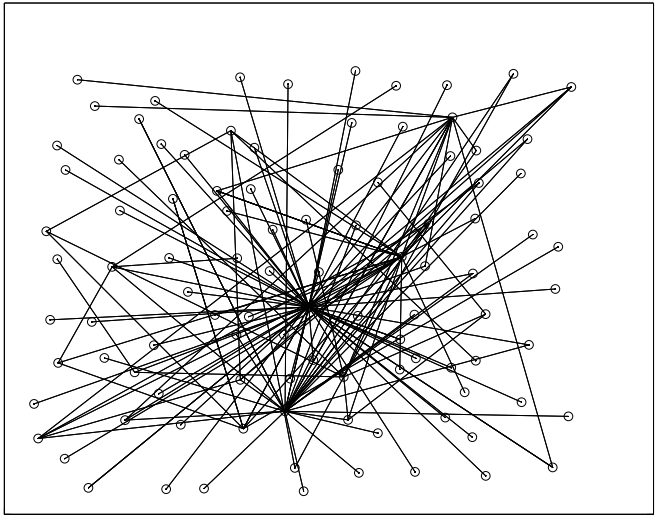}}
\caption{\label{fig:config} Small-world user network configuration.}
\end{figure}

\begin{figure}[!t]
\centerline{\epsfxsize=8.75cm\epsfbox{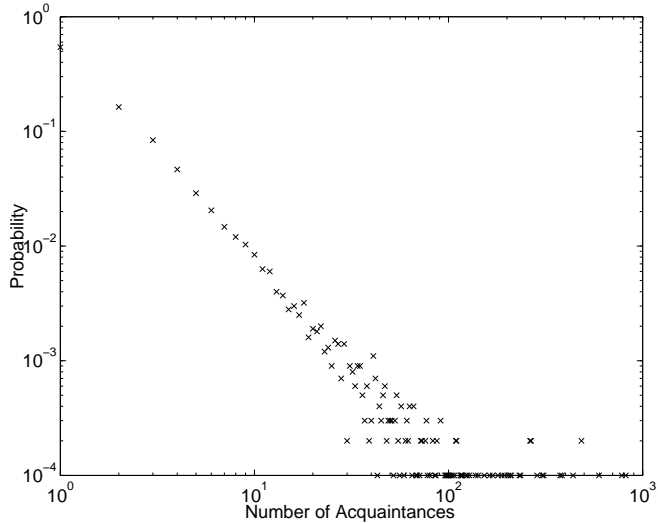}} \caption{\label{fig:powerlaw} Probability
of user $i$ having $n_i$ acquaintances versus $n_i$, showing power-law
distribution. Mean $n_i$ is 5.}
\end{figure}

The user network configuration is shown in Fig.~\ref{fig:config}, which assumes a typical small-world
configuration \cite{watts}. Specifically, each node represents a user, and a link between two
nodes indicates that these two users are acquaintances. For user $i$, the number of acquaintances
$n_i$ is a random number. It has been found that many human networks are small-world scale-free
networks, with $n_i$ typically following a power-law distribution \cite{albert}:
\begin{equation}
p(n_i)\sim n_i^{-\gamma}, \label{eq:powerlaw}
\end{equation}
where $p(n_i)$ is the probability that user $i$ has $n_i$ acquaintances and
$\gamma>0$ is the characteristic exponent. Figure \ref{fig:powerlaw} shows a power-law
distribution of $n_i$ in a small-world and scale-free user network. We clearly see that a
relatively small number of users have a large number of acquaintances. We call these users
``super users" who are responsible for a large portion of the network traffic. In a study of
long distance call traffic by Aiello \textit{et al.} \cite{aiello}, the incoming and outgoing
connections were found to follow a power-law distribution, similar to (\ref{eq:powerlaw}), and the
exponents $\gamma_{\rm in}=\gamma_{\rm out}$ was about 2.1. This clearly suggests that users do
not contribute equally to the network traffic. In the following section, we will study this effect
in detail.

\section{Simulation results}

We consider a telephone network of $N$ users. Users are located in $M$ subnetworks, each
supporting $N/M$ users.The use of subsystems is to reflect the actual case in modern telephone
systems. In a fixed telephone system, the subsystems are the central offices; in a cellular mobile
system, the subsystems are referred to as cells. Here, for simplicity, we assume that users remain
in their subnetworks for the entire period of simulation.\footnote{In the case of mobile networks,
the traffic behavior may be further complicated by the dynamics of users moving from one
subnetwork to another at different times.} Two user network configurations, namely, the
fully-connected network and scale-free network, are considered.

In a fully-connected user network, each user can call any other user with equal probability. The
effect of each user is assumed to be identical. Thus, each user has the same average arrival rate,
i.e., $\mu_i=\overline{\mu}$ and $\lambda_i=\overline{\lambda}$ for all $i$. In this way, the
classical traffic analysis ignores the effect of user network behavior on the traffic.

In a scale-free user network, each user has his own acquaintance list. A user can only call his
own acquaintances. The following two-step method is used to construct the scale-free user network.
First, the number of acquaintances $n_i$ for user $i$ is determined by a power-law distributed
random number generator. In other words, the size of the acquaintance list for each user is fixed in
this step. Next, the acquaintance lists are filled by randomly selecting acquaintances. The
relationship of acquaintance is bi-directed. If user $i$ is selected as an acquaintance of user
$j$, then user $j$ is automatically added into user $i$'s acquaintance list.
When a user is going to make a call, he randomly chooses a
receiver from his acquaintance list.

As mentioned before, the probability that a user with more acquaintances makes/receives a call is
higher. Then, the mean value of this user's inter-call time is smaller. In order to show this inequality,
we assume
\begin{equation}
\mu_i=p_0 n_i, \label{eq:mu}
\end{equation}
where $p_0$ is a constant of proportionality.

The simulation parameters are set as follows:
\begin{quote}
\vspace*{-1ex}$N= 10000, \;\;  M = 4, \;\;
\overline{n}= {\rm average}~n_i = 5$,\\
$p_0 = {1}/{500}$~call/min/acquaintance,\\
$t_m = 4~\mbox{min},$\\
$\overline{\mu} = p_0 \overline{n} = 0.01$~call/min.
\end{quote}
To ensure a fair comparison, we set $\mu = 0.01$ call/min. for the fully-connected user
network. By setting these parameters, the user behavior is fixed.

\begin{figure}[t]
\centerline{\epsfxsize=8.75cm\epsfbox{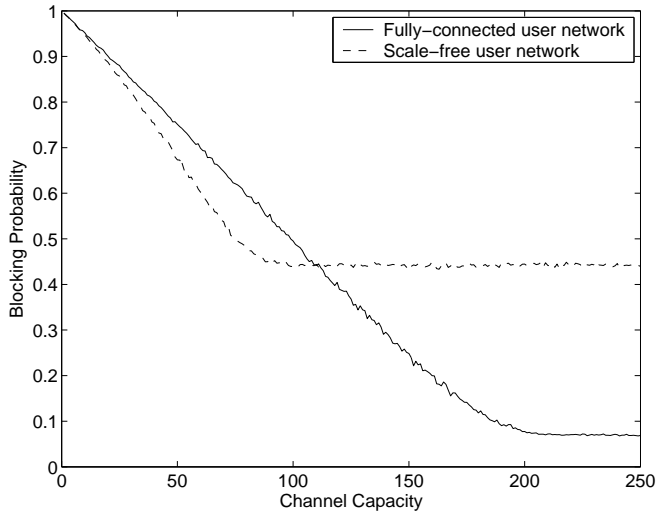}}
\caption{\label{fig:blocking_capacity} Blocking probability versus channel capacity showing
threshold effect.}
\end{figure}

Figure~\ref{fig:blocking_capacity} shows the blocking probability versus the channel
capacity.\footnote{Here the channel capacity is actually the number of channels provided in each
subnetwork.} As discussed in Section~\ref{Traffic_Analysis}, call blockings are controlled by two
factors, i.e.,  user engagement and limited channel capacity. The effect of these two factors are
clearly shown in Fig.~\ref{fig:blocking_capacity}. When the channel capacity is very limited, almost all
call arrivals are blocked. As the channel capacity increases, some of the arrived calls are
successfully set up. The call blocking probability drops. The larger the channel capacity is, the
lower the blocking probability is.  However, as the capacity reaches a certain threshold,
  the blocking probability settles to a constant value.
This clearly shows that when the channel capacity is beyond the threshold, channel capacity is no
longer a factor for call blockings and user engagement becomes the only limiting factor.
Further, the channel capacity threshold is related to the user network configuration. Our simulation (for the chosen
set of parameters) shows that  the capacity threshold for the
fully-connected user network is about 210 channels per subnetwork, and is
only about 100 channels per subnetwork for the scale-free user network. Moreover, the blocking probability
for the scale-free user network settles to about 44\%, and is much higher than that
for the fully-connected user network, which is about 7\%. The generally higher
blocking probability for the scale-free user network is caused by call concentration on a
small number of users who have a relatively large number of acquaintances.

\begin{figure}[t]
\centerline{\epsfxsize=8.75cm\epsfbox{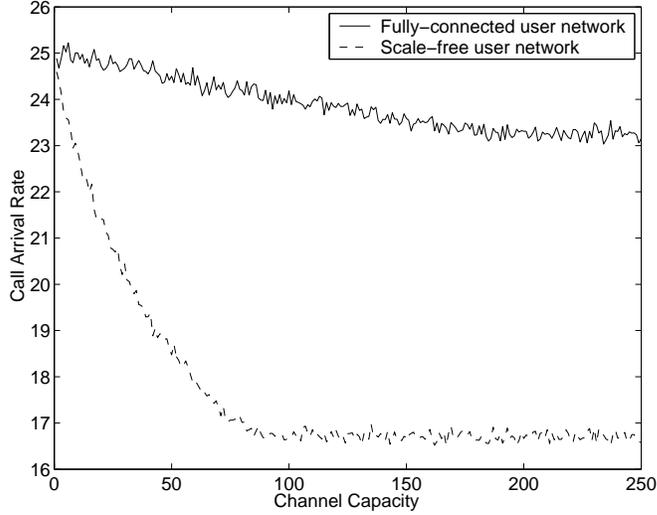}} \caption{\label{fig:allcall_capacity}
Average call arrival rate versus channel capacity.}
\end{figure}

\begin{figure}[t]
\centerline{\epsfxsize=8.75cm\epsfbox{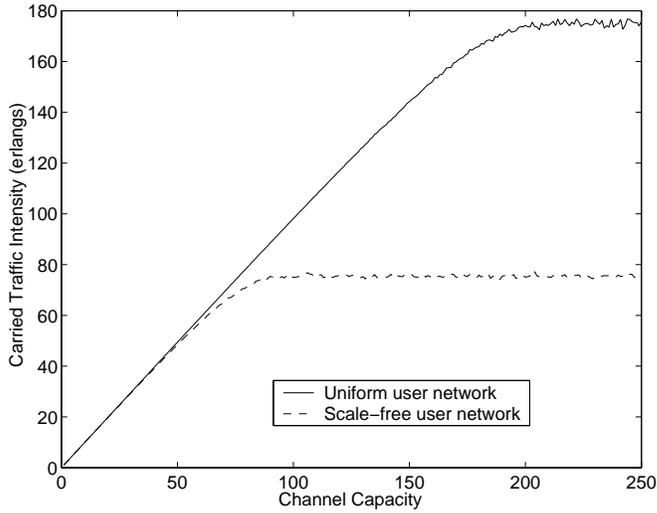}} \caption{\label{fig:traffic_capacity}
Carried traffic intensity versus channel capacity.}
\end{figure}

Figure~\ref{fig:allcall_capacity} shows the actual call arrival rate versus the channel capacity.
From this figure, we can make two main observations. First, the threshold effect exists in both user
network configurations. Before the capacity reaches the threshold, the call arrival rate decreases
as network capacity increases. When the channel capacity reaches the threshold, the call arrival
rate is almost fixed. The small fluctuations in the resulting curves are due to the randomness of
call processes in our simulation. Second, noticeable differences between the simulation results of
the two user networks are found. The call arrival rate for the scale-free user network declines more
rapidly than that for the fully-connected user network. Furthermore, the thresholds for the two user
networks are different, and the call arrival rates beyond the corresponding thresholds are also
different. For the fully-connected user network, the call arrival rate is between 23 and 24 call/min after
the threshold is reached. For the scale-free user network,
the call arrival rate is between 16 and 17 call/min after the threshold is reached.

The decrease of the call arrival rate as channel capacity increases is due to the complex calling
processes. As discussed in Section~\ref{Traffic_Analysis}, there are three typical calling
processes. The average inter-arrival times in the three cases are different. The shortest inter-arrival
time happens in Case II. The actual calling process is the combination of the three typical calling
processes. When the channel capacity is low, the channels are more likely to be fully occupied
and Case II (i.e., call blocking) is more likely to occur.
The average inter-arrival time is thus shorter, and the average arrival rate is
higher. As the channel capacity increases, the blocking probability drops. Thus, the average
inter-arrival time becomes longer, making the average arrival rate lower. When the
channel capacity reaches the threshold, the blocking probability  becomes steady, and the average call
arrival rate remains almost unchanged.

The resulting carried traffic intensities are shown in Fig.~\ref{fig:traffic_capacity}.
The carried traffic intensity is a function of the call arrival rate and blocking probability,
as in (\ref{eq:carried-traffic}).  Hence, when a drop in call arrival rate is ``over-compensated'' by a
reduction in blocking probability, the net effect is an  increase in carried traffic intensity.
This phenomenon occurs when the channel capacity is increased initially. As the channel capacity
is increased beyond the threshold, both the call arrival rate and the blocking probability arrive at constant values,
and the corresponding carried traffic intensity also becomes steady.
From the Fig.~\ref{fig:traffic_capacity}, it can be observed that
when the channel capacity is higher than 100 channels, the carried traffic
intensity for the scale-free user network stays at about 75 erlangs. When the channel capacity is
beyond 210 channels, the carried traffic intensity for the fully-connected user network remains at about 175 erlangs.

The simulated results may seem to deviate from our usual expectation. The normal way to avoid call
blockings is to increase the network capacity. But our simulation results show that in addition to
inadequate channel capacity, the user network configuration has a profound influence on call blockings.
Increasing the network capacity may not solve the problem. The user network configuration
must be considered when making telephone network planning. Our simulation also shows that the
traffic for the scale-free user network differs significantly from that for the fully-connected
user network, which is usually assumed in classical traffic analyses. For example, the channel
capacity threshold for the fully-connected user network is about 210 channels, whereas the threshold for
the scale-free user network is only about 100 channels. Because of the scale-free nature of human networks,
analyses based on a scale-free user network should reflect more realistic traffic scenarios.

\begin{figure}[!t]
\centerline{\epsfxsize=7cm\epsfbox{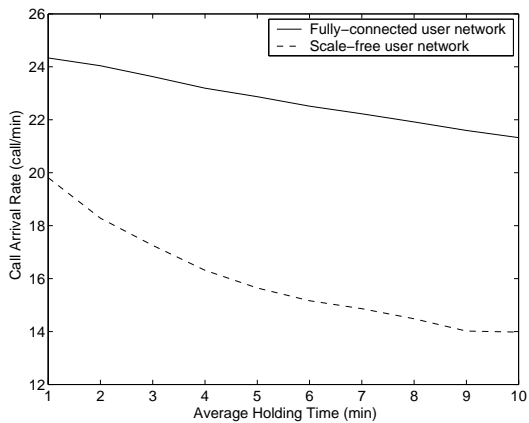}}

\centerline{\small (a)}

\centerline{\epsfxsize=7cm\epsfbox{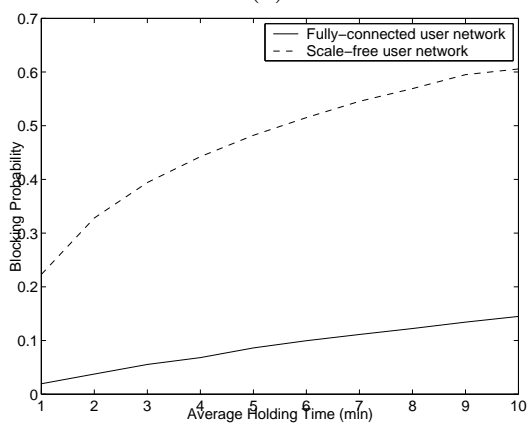}}

\centerline{\small (b)}

\centerline{\epsfxsize=7cm\epsfbox{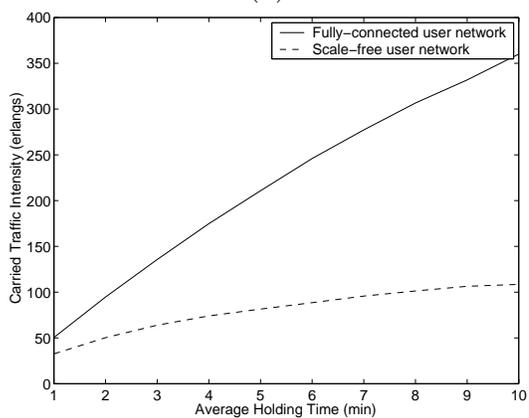}}

\centerline{\small (c)}

\caption{Effects of the average holding time on network traffic.
(a) Average call arrival rate versus average holding time;
(b) blocking probability versus average holding time;
(c) carried traffic intensity versus average holding time.} \label{fig:traffic_holdingtime}
\end{figure}

\begin{figure}[!t]
\centerline{\epsfxsize=7cm\epsfbox{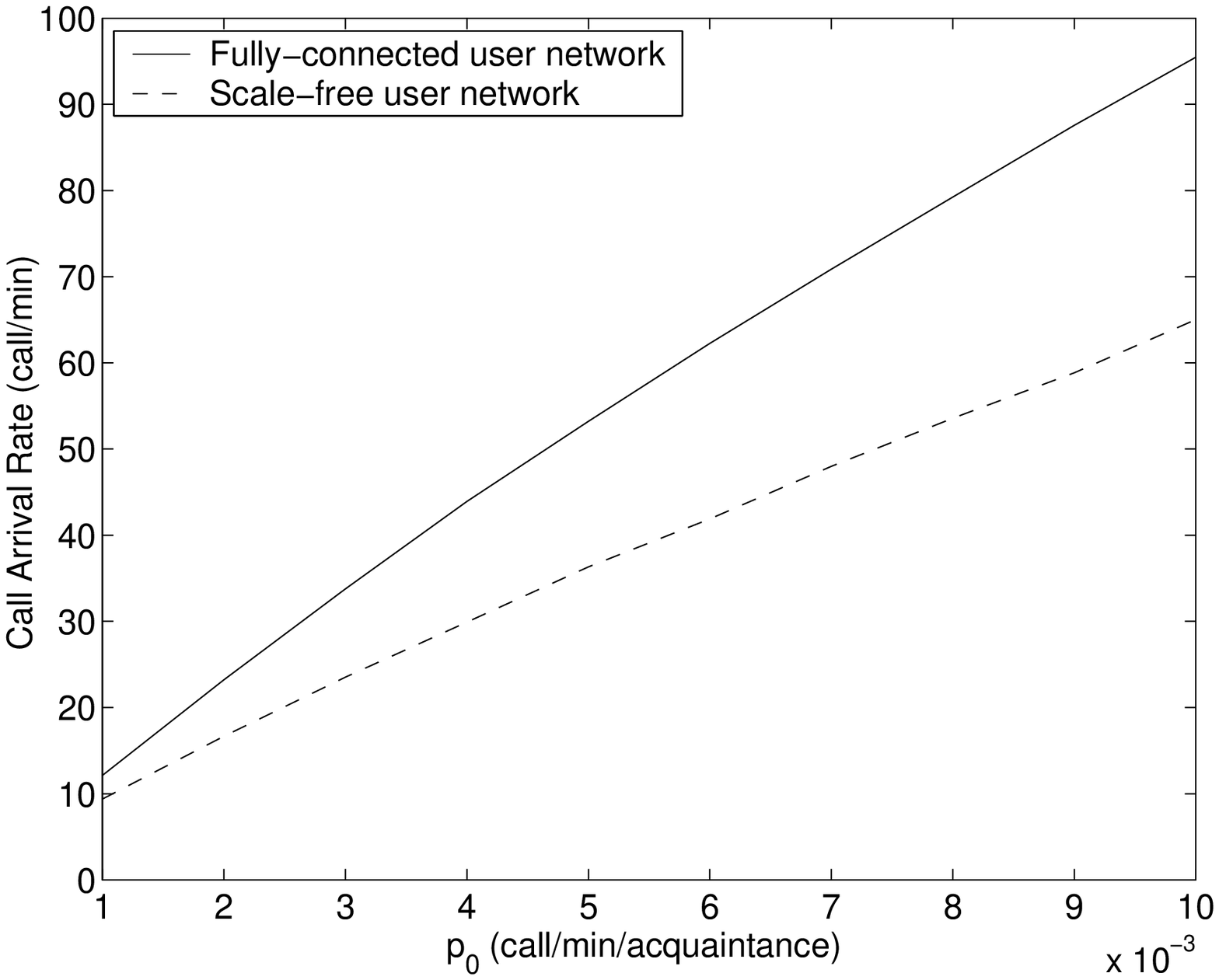}}

\centerline{\small (a)}

\centerline{\epsfxsize=7cm\epsfbox{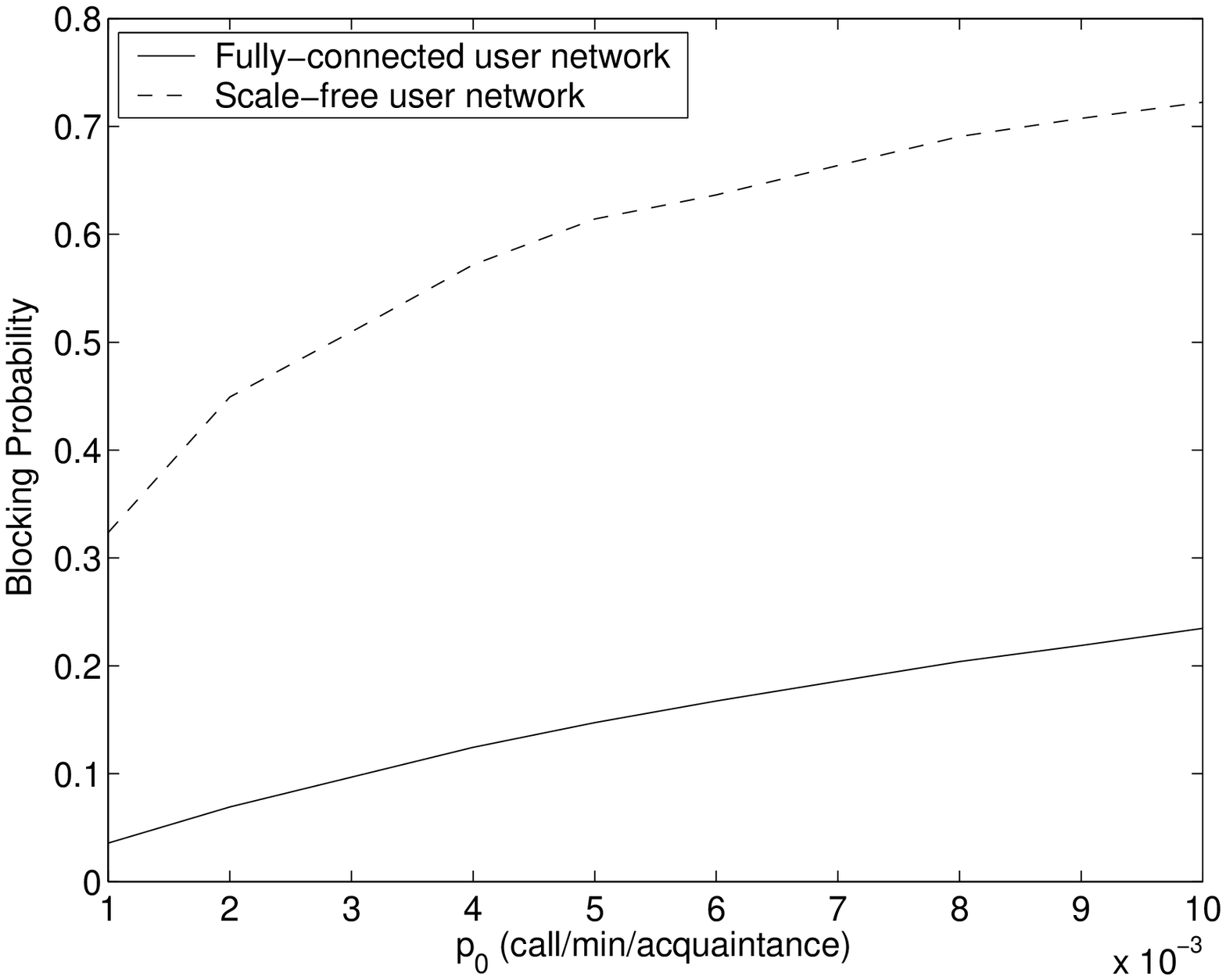}}

\centerline{\small (b)}

\centerline{\epsfxsize=7cm\epsfbox{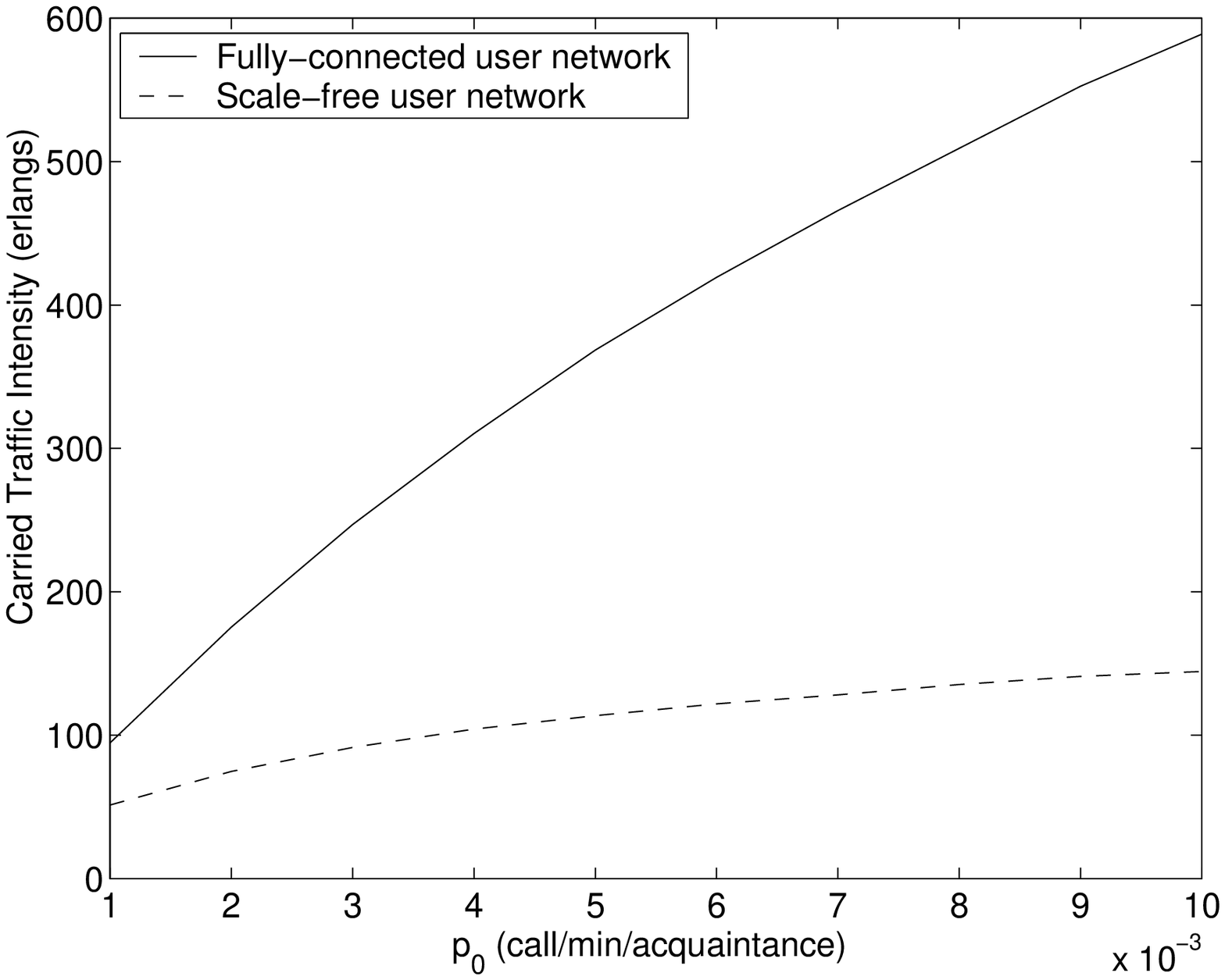}}

\centerline{\small (c)}

\caption{Effects of $p_0$ on network traffic.
(a) Average call arrival rate versus $p_0$;
(b) blocking probability versus $p_0$;
(c) carried traffic intensity versus $p_0$.} \label{fig:traffic_p0}
\end{figure}

\section{Effects of varying network parameters}
Varying network parameters may influence the traffic load in a telephone system. In this
section, we study the effects of three important parameters, namely,
the average holding time $t_m$, proportionality constant $p_0$ and
the average number of acquaintances $\overline{n}$.  In order to focus on the effects of
the network parameters,  we eliminate the effects of the channel capacity by temporarily setting it
to infinity.  For the fully-connected network,  each
user can call any other user. For a fair comparison between the fully-connected user network
and the scale-free user network,  the same set of average inter-call time ($1/\mu$) and average
holding time ($t_m$) will be used in both user networks.


First, we vary the average holding time $t_m$ in both user networks and show the simulation results
  in Fig.~\ref{fig:traffic_holdingtime}. In both
user network configurations, we observe that by increasing $t_m$, the call arrival rate decreases.
This can be reasoned as follows. For the usual Case I, since the inter-arrival time is the sum of the
holding time and the inter-call time, the inter-arrival time increases as $t_m$ increases and
hence the call arrival rate decreases.
Also, as $t_m$ increases, Case III of the calling process occurs with a
higher probability, meaning that more call attempts are cancelled without being counted as a call
arrival.
The inter-arrival time will then increase, reducing the call arrival rate.
Fig.~\ref{fig:traffic_holdingtime}~(b) plots the blocking probability versus $t_m$.
With a larger $t_m$, each call lasts longer. Hence, when an incoming call
arrives, the probability that it will be blocked is higher. Thus, the blocking probability
increases with $t_m$.
Figure~\ref{fig:traffic_holdingtime}~(c) plots the carried traffic intensity versus $t_m$.
Although there is a drop in  the average call arrival rate and an increase in the blocking probability
with $t_m$, the carried traffic intensity still grows because the increase in the average call duration
``over-compensates" the aforementioned two effects.

In the second set of simulations, we vary the proportionality constant $p_0$ in the
scale-free user network. Correspondingly, the value of $\mu$ in the fully-connected network
changes according to $\mu = 5 p_0$ to maintain the same average inter-call time ($1/\mu$) in both networks.
Figure \ref{fig:traffic_p0} shows the effect of changing $p_0$ (and hence $\mu$ in the case
of the fully-connected network).  For a larger $p_0$, the
probability of initiating a call from any user is higher. The call arrival rate thus increases and
consequently the blocking probability increases. Comparing the fully-connected user network with the scale-free user
network, the carried traffic intensity for the scale-free user network grows much slower than that for the
fully-connected user network because of the slow increase in the call arrival rate and rapid
increase in the blocking probability.

\begin{figure}[!t]
\centerline{\epsfxsize=7cm\epsfbox{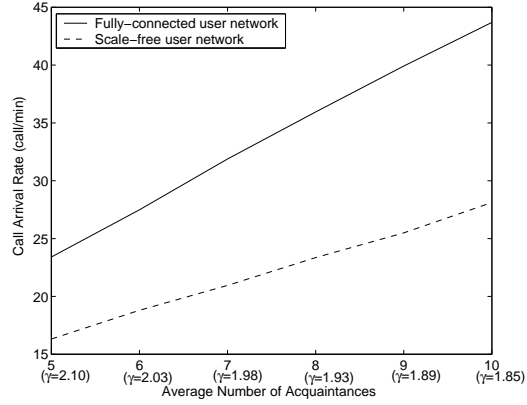}}

\centerline{\small (a)}

\centerline{\epsfxsize=7cm\epsfbox{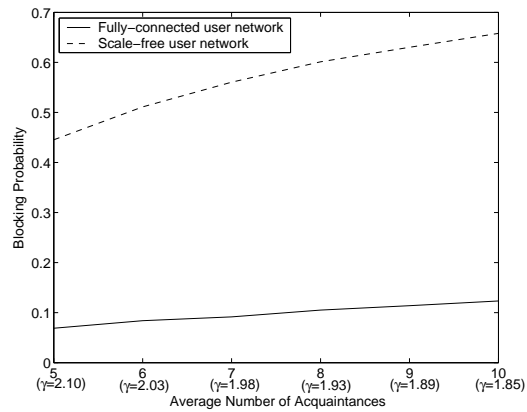}}

\centerline{\small (b)}

\centerline{\epsfxsize=7cm\epsfbox{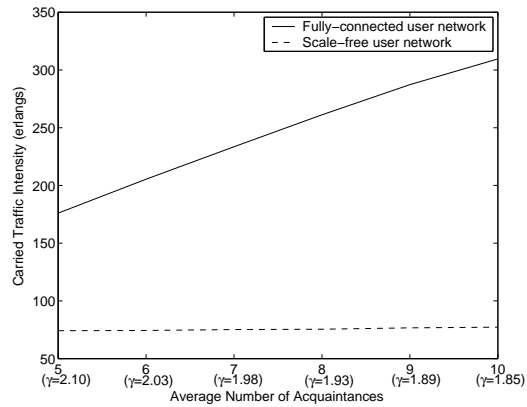}}

\centerline{\small (c)}
\caption{Effects of the average number of acquaintances on network traffic.
(a) Average call arrival rate versus average number of acquaintances;
(b) blocking probability versus average number of acquaintances;
(c) carried traffic intensity versus average number of acquaintances.}
\label{fig:traffic_numberoffriends}
\end{figure}

Finally, the effect of varying the average number of acquaintances $\overline{n}$ is shown in
Fig.~\ref{fig:traffic_numberoffriends}. For the scale-free user network, we adjust this parameter
by changing $\gamma$ of the power-law distribution. As shown in (\ref{eq:powerlaw}), a smaller
$\gamma$ corresponds to a gentler slope of the power-law distribution, which means that more users
have a large number of acquaintances. Hence, $\overline{n}$ increases as $\gamma$ decreases.
Correspondingly, in the fully-connected user network, we change the value of $\mu$ according to
$\mu = \overline{n}/500$ to maintain the same average inter-call time ($1/\mu$) in both user
networks.

As shown in (\ref{eq:mu}), calls arrive more frequently when $\overline{n}$ (and $\mu$) increases.
Thus, the average
call arrival rate is found to increase with $\overline{n}$,
as shown in Fig.~\ref{fig:traffic_numberoffriends}~(a).
The increase in call arrivals causes an increase in the blocking probability,
 as shown in Fig.~\ref{fig:traffic_numberoffriends}~(b). In
Fig.~\ref{fig:traffic_numberoffriends}~(c),
the carried traffic intensity is plotted versus the average number of acquaintances.
For the fully-connected user network, the increase in the call arrival rate
``overshadows'' the increase in the blocking probability, causing a net rise
in the carried traffic intensity.
For the scale-free user network, the effect of an increase in the call arrival rate
is ``balanced out'' by an increase in the blocking probability, causing no apparent
change in the carried traffic intensity.

As a final and general remark, compared with the telephone system
with a fully-connected user network, the system with a scale-free user network has a lower average
call arrival rate, a (much) higher blocking probability, and a lower carried traffic intensity.

\section{Conclusions}
This paper studies the telephone network traffic from a scale-free user network perspective. Two
major factors, namely, the channel capacity and user network configuration, are identified as
being pivotal to call blockings.  Our simulation results show that the network traffic assuming a
scale-free user network differs substantially from the traffic assuming a conventional
fully-connected user network. For the scale-free user network, the traffic load arises mainly from
a small number of users who have a relatively large number of acquaintances. This concentration
causes a higher blocking probability. At the same time, the majority of users, who have a few
acquaintances, contribute much less to the traffic load. In this paper we have also studied the
effects of different network parameters on the calling process. Our final conclusions are that
telephone network traffic is greatly influenced by user behavior, and that beyond a certain
capacity threshold call blockings are not likely to be reduced by increasing network capacity
(adding extra resources or intensifying investments) which would have been the usual expectation.
Thus, a clear, though obvious, lesson to be learnt from this traffic analysis is that any strategy
for altering the traffic in any manner must take into account the scale-free property of user
networks.

\end{document}